\documentstyle[sprocl,epsfig]{article}

\input{paperdef}
\def\al{\alpha}

\def\be{\begin{equation}}
\def\ee{\end{equation}}
\def\bea{\begin{eqnarray}}
\def\eea{\end{eqnarray}}

\begin{document}

\thispagestyle{empty}
\setcounter{page}{0}
\def\thefootnote{\fnsymbol{footnote}}

\begin{flushright}
CERN--PH--TH/2004--158\\
hep-ph/0408340 \\
\end{flushright}

\vspace{1cm}

\begin{center}

{\large\sc {\bf MSSM Higgs Physics: Theoretical Developments}}
\footnote{talk given at the ``International Conference on
  Linear Colliders'',
April 2004, Paris, \mbox{}~~~~~~France}

\vspace{1cm}

{\sc S.~Heinemeyer$^{\,}$%
\footnote{
email: Sven.Heinemeyer@cern.ch
}%
}

\vspace*{0.5cm}

CERN TH Division, Dept.\ of Physics, CH~1211 Geneva 23, Switzerland

\end{center}

\vspace*{1cm}

\begin{abstract}
The corrections to the MSSM Higgs boson masses and couplings performed
in this  
millennium are briefly reviewed. For the lightest MSSM Higgs boson
mass, $\mh$, we list the current status of the intrinsic uncertainties
(due to unknown higher-order corrections) and the parametric
uncertainties (due to the imperfect experimental knowledge of the
input parameters). The need for high-precision calculations in the
MSSM Higgs boson sector is exemplified in a realistic example.
\end{abstract}

\def\thefootnote{\arabic{footnote}}
\setcounter{footnote}{0}

\newpage

\title{MSSM HIGGS PHYSICS: THEORETICAL DEVELOPMENTS\\}

\author{ SVEN HEINEMEYER }

\address{CERN TH Division, Dept.\ of Physics, CH~1211 Geneva 23, Switzerland}


\maketitle\abstracts{
The corrections to the MSSM Higgs boson masses and couplings performed
in this  
millennium are briefly reviewed. For the lightest MSSM Higgs boson
mass, $\mh$, we list the current status of the intrinsic uncertainties
(due to unknown higher-order corrections) and the parametric
uncertainties (due to the imperfect experimental knowledge of the
input parameters). The need for high-precision calculations in the
MSSM Higgs boson sector is exemplified in a realistic example.
}


\section{Introduction}

Disentangling the mechanism that controls electroweak symmetry
breaking is one 
of the main tasks of the current and next generation of colliders. The
prime candidates are the Higgs mechanism within the Standard Model (SM) or
within the Minimal Supersymmetric Standard Model (MSSM)~%
\footnote{
We assume all parameters in the MSSM to be real.
}%
. Contrary to
the SM, two Higgs doublets are required in the MSSM, resulting in five
physical Higgs bosons: the light and heavy CP-even $h$ and $H$, the
CP-odd $A$, and the charged Higgs bosons $H^\pm$.
The Higgs sector of the MSSM can be expressed at lowest
order in terms of $\MZ$, $\MA$, and $\tb = v_2/v_1$, the ratio of the
two vacuum expectation values. 
The MSSM Higgs boson sector receives
large corrections from SUSY-breaking effects in the Yukawa sector of the
theory. The leading one-loop correction to the lightest MSSM Higgs
boson mass, $\mh$, is proportional to $\mt^4$. For
instance, the leading logarithmic one-loop term (for vanishing mixing
between the scalar tops) reads~\cite{mhiggs1l} 
\BE
\De \mh^2 = \frac{3 G_F \mt^4}{\wz\, \pi^2\,\SQb}
          \ln \KL \frac{\mste \mstz}{\mt^2} \KR~.
\EE
Here $m_{\tilde t_{1,2}}$ denote the masses of the scalar top quarks.
Corrections of this kind have dramatic effects on the predicted value of
$\mh$ and many other observables in the MSSM Higgs sector. The one-loop
corrections, which are known completely, can shift $\mh$ by 50--100\%. 
Since this shift is related to
effects from a part of the theory that does not enter at tree level,
corrections even of this size do not invalidate the perturbative
treatment.

The large corrections to $\mh$ have to be compared with the
anticipated accuracy at future colliders. At the LHC $\mh$ can
possibly be measured down to $\de\mh^{\rm exp, LHC} \approx 200 \mev$.
A future $e^+e^-$ linear collider (LC) can even go down to 
$\de\mh^{\rm exp,LC} \approx 50 \mev$. 

The effect of these large corrections is two-fold. On the one hand,
reliable investigations (comparisons of experimental results with
theoretical predictions) require multi-loop higher-order
corrections. On the other hand, the strong dependence of $\mh$ on the
whole SUSY spectrum, especially on the scalar top sector, makes $\mh$
an extremely powerful precision observable. $\mh$ can be used to obtain
information on otherwise unknown parameters.
This allows in particular to obtain
indirect information on the mixing in the scalar top sector, which is
very important for fits of the SUSY Lagrangian to (prospective)
experimental data~\cite{sfittino}.


\section{Recent Calculations In The MSSM Higgs Sector}

Concerning the two-loop
effects, their computation is quite advanced, see \cite{mhiggsAEC} and
references therein. They include the strong corrections
at \order{\al_t\als}, and Yukawa corrections, \order{\al_t^2},
to the dominant one-loop \order{\al_t} term, as well as the strong
corrections from the bottom/sbottom sector at \order{\al_b\als} in the
limit of $\tb \to \infty$. %
\footnote{For the result for arbitrary $\tb$ see \cite{mhiggsFD2}.
}%
~For the $b/\Sbot$~sector
corrections also an all-order resummation of the $\tb$-enhanced terms,
\order{\al_b(\als\tb)^n}, is known~\cite{deltamb}.
Most recently the \order{\al_t \al_b} and \order{\al_b^2} corrections
have been derived~\cite{mhiggsEP5}~%
\footnote{
Leading corrections in the MSSM with non-minimal flavor violation 
have recently been obtained in~\cite{mhiggsNMFV}.
}%
.~Finally a ``full'' two-loop
effective potential 
calculation (including even the momentum dependence for the leading
pieces) has been published~\cite{fullEP2l}. However, the latter results
have been obtained using a certain renormalization in which all
quantities, including SM gauge boson masses and couplings, are \drbar\
parameters. This makes them not applicable in the Feynman-diagrammatic
(FD) approach using the on-shell renormalization scheme, see \refse{sec:need}.


\section{Intrinsic Uncertainties}

The current intrinsic error (due to unknown higher-order corrections) 
consists of four different pieces:\\
$-$ missing momentum-independent two-loop corrections: By varying the
renormalization scale at the one-loop level, these two-loop
uncertainties can be estimated to be 
$\pm 1.5 \gev$.~\cite{feynhiggs1.2}~%
\footnote{We do not consider here the
``full'' two-loop effective potential calculation presented
in~\cite{fullEP2l} for the reasons outlined above.
}%
\newline
$-$ missing momentum-dependent two-loop corrections: since at the
one-loop level the momentum corrections are below the level of 
$2 \gev$, it can be estimated that they stay below 
$\pm 0.5 \gev$~\cite{mhiggsAEC,mhiggsWN}.\\
$-$ missing 3/4-loop corrections from the $t/\Stop$~sector: by
applying three different methods (changing the renormalization scheme 
at the two-loop level; direct evaluation of the leading terms in a
simplified approximation; numerical iterative solution of the
renormalization group equations) these corrections have been estimated
to be at about $\pm 1.5 \gev$ (see \cite{mhiggsAEC} and
references therein). \\
$-$ missing 3/4-loop corrections from the $b/\Sbot$~sector:
the corrections from the $b/\Sbot$~sector can be large if both, $\mu$
and $\tb$ are sufficiently large. For $\mu > 0$ it can been
shown~\cite{mhiggsFD2} that 
the QCD two-loop corrections give already an extremely precise result,
provided that the resummation of $(\als\tb)^n$ terms~\cite{deltamb} is
taken into 
account. On the other hand, for $\mu < 0$ the 3-loop
corrections can be up to $\pm 3 \gev$~\cite{mhiggsFD2}. Since the
results for $\amu$ favor a positive $\mu$ we do not
consider this possibility here.\\
The current intrinsic error can thus be estimated to
be~\cite{mhiggsAEC}
\BE
\de\mh^{\rm intr,today} \approx 3 \gev~,
\EE
depending in detail on the investigated point in the MSSM parameter
space. 

If the full two-loop calculation (in an FD suitable renormalization)
as well as the leading 3-loop (and possibly the very leading 4-loop)
corrections are available, the intrinsic error could be reduced to
less than about $\pm 0.5 \gev$. This seems to be possible within the
next 5--10 years.


\section{Parametric Uncertainties}

The currently induced error by $\MW$ and $\mb$ are already almost
negligible, and will be irrelevant with the future precision of these
input parameters~\cite{deltamt}. On the other hand, $\mt$ and $\als$ play a
non-negligible role. Currently we have~\cite{deltamt} (for a recent
reevaluation leading to similar results, see \cite{mhiggsWN})
\BEA
\de\mt^{\rm exp,today} \approx 4.3 \gev 
&\Rightarrow&
\de\mh^{{\rm para,}\mt} \approx 4 \gev \\
\de\al^{\rm exp,today} \approx 0.002
&\Rightarrow&
\de\mh^{{\rm para,}\als} \approx 0.3 \gev
\EEA
From the LC one can hope to achieve in the future
\BEA
\de\mt^{\rm exp,future} \approx 0.1 \gev 
&\Rightarrow&
\de\mh^{{\rm para,}\mt} \approx 0.1 \gev \\
\de\als^{\rm exp,future} \lsim 0.001
&\Rightarrow&
\de\mh^{{\rm para,}\als} \approx 0.1 \gev
\EEA

The error induced by the experimental uncertainties of the SUSY
parameters is very difficult to estimate. It will depend substantially
on the values of the parameters realized in nature. For most of the
MSSM parameter space no analysis of the anticipated experimental
errors is available yet. 


\section{The Need For Precision}
\label{sec:need}

\begin{figure}[htb!]
\begin{center}
\epsfig{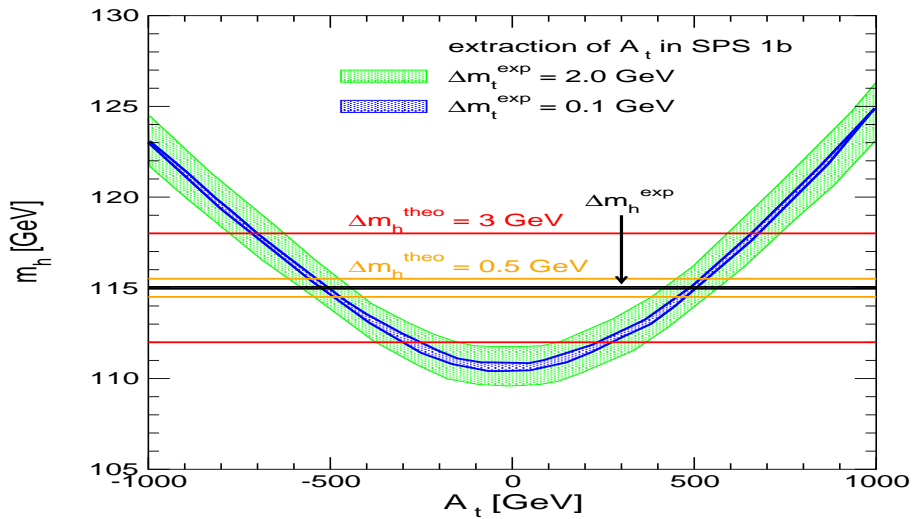}
\vspace{-2em}
\caption{The dependence of $\mh$ within the SPS~1b scenario is shown
as a function of $\At$. The band reflects the effects of parametric
uncertainties of the MSSM parameters (including prospective
experimental errors, see text). The light shaded (green)
area corresponds to $\de\mt^{\rm exp} = 2 \gev$, the dark shaded
(blue) area to $\de\mt^{\rm exp} = 0.1 \gev$ (see text). The possible
experimental measurement of $\mh$ is shown (including the error). Two
further bands are shown, demonstrating the effect of an intrinsic
error of $3 \gev$ (today) and $0.5 \gev$ (future).
}
\label{fig:need}
\vspace{-1em}
\end{center}
\end{figure}

We present one example emphasizing the need for a drastic improvement
in the intrinsic error of $\mh$ (assuming that the experimental and
parametric error will be well enough under control, see above). The
evaluation of $\mh$ is based on the code
\fh~\cite{mhiggsCPXFD1l,mhiggslong,feynhiggs}. We consider the SPS~1b
benchmark scenario~\cite{sps}, where it will be very
challenging to measure the trilinear Higgs-stop coupling, $\At$. In
\reffi{fig:need} we show the dependence of $\mh$ on $\At$ in this
scenario~\cite{deltamt}. The band reflects the effects of parametric
%
%
uncertainties of the MSSM parameters (including prospective
experimental errors, see also \cite{deltamt,deschi}). The light shaded
(green) area corresponds to $\de\mt^{\rm exp} = 2 \gev$, the dark shaded
(blue) area to $\de\mt^{\rm exp} = 0.1 \gev$. The possible
experimental measurement of $\mh$ is shown (including the error). Two
further horizontal bands are shown, demonstrating the effect of an intrinsic
error of $3 \gev$ (today) and $0.5 \gev$ (future). From the
intersection of the experimental $\mh$ determination, including the
intrinsic error, with the SPS~1b band allows to determine $\At$
indirectly (up to a sign ambiguity, see \cite{deschi} for a
determination of its sign). Besides the need for a precise $\mt$
measurement, it becomes obvious that with the current intrinsic $\mh$
uncertainty hardly any bound could be set. If, on the other hand, a
reduction of the intrinsic error down to $\sim 0.5 \gev$ could be
performed, determination of $\At$ better than $\sim 10\%$ seems feasible.


\end{document}